\documentclass[acmsmall]{acmart} 
 \usepackage{amsmath,amssymb,amsfonts,bm}
\usepackage{graphicx}
\usepackage{textcomp}
\usepackage{xcolor}
\usepackage{float}
\usepackage{array}
\usepackage{tabularx}
\usepackage{array}
\usepackage{balance}
\usepackage{marvosym}
\usepackage{multirow}
\usepackage{array}
\usepackage{enumitem}
\usepackage{booktabs}
\usepackage{algpseudocode}
\usepackage[ruled]{algorithm2e} 
\usepackage{xcolor}
\usepackage{pifont}
\usepackage{url}
\usepackage{algpseudocode}
\usepackage{subfig}
\usepackage[ruled]{algorithm2e}
\usepackage{amsmath,bm}
\usepackage{makecell}
\AtBeginDocument{%
  \providecommand\BibTeX{{%
    \normalfont B\kern-0.5em{\scshape i\kern-0.25em b}\kern-0.8em\TeX}}}

\setcopyright{acmlicensed}
\acmJournal{TOIS}
\acmYear{2023} \acmVolume{1} \acmNumber{1} \acmArticle{1} \acmMonth{1} \acmPrice{15.00}\acmDOI{XXXXXXX.XXXXXXX}

\begin{document}

\title{Efficient On-Device Session-Based Recommendation}

\author{Xin Xia}
\affiliation{%
	\institution{The University of Queensland}
	\city{Brisbane}
	\country{Australia}}
\email{x.xia@uq.edu.au}

\author{Junliang Yu}
\affiliation{%
	\institution{The University of Queensland}	
	\city{Brisbane}
	\country{Australia}}
\email{jl.yu@uq.edu.au}

\author{Qinyong Wang}
\affiliation{%
	\institution{Baidu Inc.}	
	\city{Beijing}
	\country{China}}
\email{wangqinyong@baidu.com}

\author{Chaoqun Yang}
\affiliation{%
	\institution{Griffith University}	
	\city{Gold Coast}
	\country{Australia}}
\email{chaoqun.yang@griffith.edu.au}

\author{Quoc Viet Hung Nguyen}
\affiliation{%
	\institution{Griffith University}	
	\city{Gold Coast}
	\country{Australia}}
\email{henry.nguyen@griffith.edu.au}

\author{Hongzhi Yin}
\authornote{Corresponding author.}
\affiliation{%
	\institution{The University of Queensland}
	\city{Brisbane}
	\country{Australia}}
\email{db.hongzhi@gmail.com}

\thanks{This work was supported by the Australian Research Council Future Fellowship (Grant No. FT210100624), the Discovery Project (Grant No. DP190101985),  and the Discovery Early Career Research Award (Grant No. DE200101465).}

\begin{abstract}
On-device session-based recommendation systems have been achieving increasing attention on account of the low energy/resource consumption and privacy protection while providing promising recommendation performance. To fit the powerful neural session-based recommendation models in resource-constrained mobile devices, tensor-train decomposition and its variants have been widely applied to reduce memory footprint by decomposing the embedding table into smaller tensors, showing great potential in compressing recommendation models. However, these model compression techniques significantly increase the local inference time due to the complex process of generating index lists and a series of tensor multiplications to form item embeddings, and the resultant on-device recommender fails to provide real-time response and recommendation. To improve the online recommendation efficiency, we propose to learn compositional encoding-based compact item representations. Specifically, each item is represented by a compositional code that consists of several codewords, and we learn embedding vectors to represent each codeword instead of each item. Then the composition of the codeword embedding vectors from different embedding matrices (i.e., codebooks) forms the item embedding. Since the size of codebooks can be extremely small, the recommender model is thus able to fit in resource-constrained devices and meanwhile can save the codebooks for fast local inference. Besides, to prevent the loss of model capacity caused by compression, we propose a bidirectional self-supervised knowledge distillation framework. Extensive experimental results on two benchmark datasets demonstrate that compared with existing methods, the proposed on-device recommender not only achieves an 8x inference speedup with a large compression ratio but also shows superior recommendation performance. The code is released at \url{https://github.com/xiaxin1998/EODRec}.
  \end{abstract}




\maketitle

\section{Introduction}
Recently, session-based recommender systems~\cite{liu2018stamp, shani2005mdp, wu2019session, chen2020sequence} are emerging, which aim to recommend a ranked list of items that the target user is interested in based on his/her current session context.  
Meanwhile, of all the shopping channels available to customers, mobile commerce is taking the lead \footnote{https://www.shopify.com/au/enterprise/mobile-commerce-future-trends}.
{Currently, most session-based recommendation systems are built upon a cloud-based paradigm, where the session-based recommender models are trained and deployed on powerful cloud servers \cite{disabato2020incremental, dhar2021survey}}. When a recommendation request is made by a user via a  mobile device, the current session information will be sent to the on-cloud session-based recommender models to generate recommendation results. Then, the recommendation results will be returned to the mobile device for display. However, in reality, this  recommendation paradigm heavily relies on high-quality wireless network connectivity and increasingly raises public privacy concerns. 

\par
To better protect user privacy and reduce communication costs, on-device machine learning \cite{dhar2021survey} that aims to develop and deploy machine learning models on resource-constrained devices has been attracting more and more attention. In this learning paradigm, the machine learning model is first trained on the cloud and then downloaded and deployed on local devices such as smartphones. When using the model, users no longer need to upload their sensitive data (e.g., various context information) to central servers and, therefore, can enjoy privacy-preserving and low-latency services that are orders of magnitude faster than server-side models \cite{lee2019device}. Since the principle of on-device machine learning is well-aligned with the need for low-cost and privacy-preserving recommender systems, there has been a growing trend towards on-device recommendation \cite{han2021deeprec,wang2020next,ochiai2019real,changmai2019device,chen2021learning,yin2021tt}.
However,  it is non-trivial to deploy the powerful recommender model trained on the cloud to resource-constrained mobile devices, which are confronted with the following two fundamental challenges.\par
The first challenge is \textbf{how to reduce the size of on-cloud trained models to fit in resource-constrained devices without compromising recommendation performance}. To address this challenge,  a typical approach is to employ knowledge distillation \cite{hinton2015distilling} that first trains a \textit{teacher} model on the cloud by fully utilizing the abundant resources and then transfers the teacher's knowledge to a lightweight \textit{student} model.  To retain the original recommendation performance,  even the \textit{student} model cannot be small enough to fit in resource-constrained mobile devices without compression.  But the classical model compression techniques such as pruning \cite{srinivas2015data}, down-sizing \cite{hinton2015distilling}, and  parameter sharing \cite{plummer2020shapeshifter}  are not so effective in compressing the recommender models. Unlike the vision and language models \cite{han2015deep}, which are usually over-parameterized with a very deep network structure, the recommendation model only contains several intermediate layers that account for a small portion of learnable parameters. Instead, the item embedding table is the one that accounts for the vast majority of memory footprint \cite{wu2020saec}. Recent practices of on-device recommendation \cite{wang2020next,sun2020generic} mostly adopt decomposition techniques such as tensor-train decomposition (TTD) \cite{oseledets2011tensor} to reconstruct the embedding table with a series of matrix multiplication. However, to avoid a drastic recommendation performance drop, they can only compress the embedding table with a small compression rate (e.g., 4$\sim$8x), which is far from meeting the resource-constrained conditions on mobile devices. In addition, recent studies show that models with similar structures (e.g., encoder-decoder) are easier to transfer knowledge \cite{chen2021knowledge}, while the tensor decomposition widens the structural gap between the teacher and the student.


The second challenge is \textbf{how to provide real-time response and recommendation}. When a user interacts with new items in the current session, the recommendation system should provide a response by updating recommendations immediately.  The response speed to generate new recommendations is crucial for user experience in mobile e-commerce. However, the widely used recommendation model compression techniques (tensor-train decomposition and its variants) significantly delay the on-device model inference due to their complex process of generating index lists and a series of multiplications to form item embeddings, and the resultant on-device recommenders fail to provide real-time response and recommendation. 

In our previous work~\cite{xia2022device}, we proposed semi-tensor decomposition and self-supervised knowledge distillation techniques to address the first challenge, based on which an ultra-compact and effective on-device recommendation model \textbf{OD-Rec} was developed. Specifically, given the dilemma that the small TT-rank in tensor-train decomposition
leads to under-expressive embedding approximations, whereas the larger TT-rank sacrifices the model efficiency~\cite{novikov2015tensorizing, oseledets2011tensor}, we introduced the semi-tensor product (STP) operation~\cite{zhao2021semi} to tensor-train decomposition for the extreme compression of the embedding table. This operation endows tensor-train decomposition with the flexibility to perform multiplication between tensors with inconsistent ranks, considering both effectiveness and efficiency. With semi-tensor decomposition to compress the embedding table, we achieved an extremely compact model of 30x smaller size than the original model. In addition, we proposed recombining the embeddings learned by the teacher and the student to perform self-supervised learning, which can further distill the essential information lying in the teacher’s embeddings to compensate for the accuracy loss caused by the extremely high compression. However, the second challenge has not been addressed in~\cite{xia2022device}.  As a variant of tensor-train decomposition, the process of reconstructing item embedding in our semi-tensor decomposition is also very complex, which significantly slows down the online model inference to generate recommendations. An index list that is used to map the rows and columns in the factorized tensors should be first calculated, and it takes more time to do a series of multiplication between the mapped vectors.


In light of this,  we extend our previous work~\cite{xia2022device} by developing a new compositional code-based compression method to replace the previous semi-tensor decomposition.  In the new method, we propose representing each item by a unique $M$-dimensional code consisting of $M$  discrete codewords, where $M$ is much smaller than the original embedding dimension.
There are $M$ codebooks, each containing $K$ codeword vectors (i.e., basic vectors). We learn embedding vectors for each codeword rather than each item and then simply sum up the corresponding codeword vectors to construct item embedding.  For example, given a compositional code $C_i = (2,3,5,1)$ for item $i$, and  $4$ codebooks $E_1, E_2, E_3, E_4$, then the item embedding of $i$ is computed as
\begin{equation}
E\left(C_{i}\right)=\sum_{t=1}^{4} E_{t}\left(C_{i}^{t}\right),
\end{equation}
where $E_{t}(C_{i}^{t})$ is $C_{i}^{t}$-th codevector in codebook $E_t$. To learn effective item compositional embeddings, we minimize the difference between them and the well-trained uncompressed item embeddings. Meanwhile, the recommendation task is jointly optimized with this regularization term to ensure a smooth knowledge transfer. After convergence, the well-trained codebooks and codes are saved to be used in the on-device recommender. Therefore, on the resource-constrained devices, the total memory footprint for items is of size $|V|M + MKN$, where $N$ denote the dimension of a codeword vector. We can adjust the value of $M$ to compress the item embedding table to any degree. In addition, our previous knowledge distillation framework only transfers the teacher's knowledge to the student based on the assumption that the teacher always outperforms the student. However, our experiments in~\cite{xia2022device} demonstrated that the lightweight student model achieves better recommendation performance on some metrics, which is  consistent with the findings in~\cite{kweon2021bidirectional}. Inspired by this observation, we propose a bidirectional knowledge distillation framework that also enables the knowledge transfer from the student to the teacher. It means that both the teacher and the student are iteratively updated in the framework. Particularly, we find that the student and the teacher trained under this new framework are superior to those trained under our previous framework. \par

To summarize, our major new contributions are listed as follows:
\begin{itemize}
	\item We make the exploration of ultra-compact efficient on-device session-based recommendation by integrating discrete compositional code learning into recommender systems to compress item embedding table. 
	\item We propose a bidirectional knowledge distillation framework that enables the teacher and the student to mutually enhance each other during the learning process.  
	\item We conduct extensive experiments to evaluate the online inference efficiency, recommendation accuracy, and model compression ratio. The experimental results on two benchmark datasets demonstrate that the proposed new method completely outperforms the previous ones, showing an 8x speedup in model inference with smaller model size and higher recommendation accuracy.
\end{itemize}
The remainder of the paper is organized as follows. In Section 2, the related work is reviewed in
detail, followed by the description of the preliminaries in Section 3. In Section 4, the detail of the proposed compression method is presented. Then, we show how to use self-supervised knowledge distillation to prevent the loss of model capacity in Section 5. Extensive experiments are conducted and analyzed to verify the efficacy of the proposed model in Section 6. At last, we conclude our work and discuss the future work in Section 7.

\section{Related Work}
In this section, we briefly review the literature on four related topics: on-device recommendation, session-based recommendation, knowledge distillation for recommendation, and compositional code learning. The former two are the scenarios where our method is applied, and the latter two are techniques that our method involves.

\subsection{On-Device Recommendation}
On-device machine learning \cite{dhar2021survey} has been drawing increasing attention in many fields due to its advantages in local inference and privacy protection. The fundamental challenge of on-device learning is how to reduce the size of conventional models so as to adapt to resource-constrained devices. The existing research \cite{chen2021learning, changmai2019device, han2021deeprec, ochiai2019real} mainly adopts model compressing techniques like pruning \cite{han2015deep, srinivas2015data}, quantization \cite{gong2014compressing}, and low-rank factorization \cite{novikov2015tensorizing, oseledets2011tensor} to achieve this goal. Recently, on-device recommender systems emerged in response to users' needs for low latency recommendation and have demonstrated desired performance with tiny model sizes. Specifically, Chen \textit{et al.} \cite{chen2021learning} proposed to learn elastic item embeddings to allow recommendation models to be customized for arbitrary device-specific memory constraints without retraining. WIME \textit{et al.} \cite{changmai2019device} proposes to use user's real-time intention to generate on-device recommendations where a sequence and context aware algorithm was used to embed user intention and a neighborhood searching method followed by a sequence matching algorithm was used to do prediction. DeepRec \cite{han2021deeprec} uses model pruning and embedding sparsification techniques to enable model training with a small memory budget and fine-tunes the model using local user data, reaching comparable performance in sequential recommendation with a 10x parameter reduction. Wang \textit{et al.} \cite{wang2020next} proposed LLRec which is for next POI recommendation model and is based on tensor-train factorization. TT-Rec \cite{yin2021tt} also applies tensor-train decomposition to the embedding layer for model compression and further improves model performance with a sampled Gaussian distribution for the weight initialization of the tensor cores. Despite the effectiveness of these methods, they are not designed for session-based recommendation, and many of them are still with a large number of parameters which may overload small devices. 

\subsection{Session-Based Recommendation}
The early session-based recommendation models \cite{shani2005mdp, yin2016spatio} are based on Markov Chain and mainly focus on modeling temporal order between items. In the deep learning era, neural networks like RNNs \cite{hidasi2015session, chen2020sequence,zhang2018discrete} are tailored to model session data and handle the temporal shifts within sessions. The follow-up models such as NARM \cite{li2017neural} and STAMP \cite{liu2018stamp} employ attention mechanisms to prioritize items when profiling users' main interests. Pan \textit{et al.} \cite{pan2020rethinking} proposed to rethink item importances in the session-based recommendation and estimate the importance of items in a session by employing an Importance Extracting Module based on a modified self-attention mechanism. Recently, graph-based methods \cite{wu2019session, qiu2020gag, 9101653} have demonstrated state-of-the-art performance, which constructs various session graphs to model item transitions. Specifically, SR-GNN \cite{wu2019session} constructs session graphs for every session and designs a gated graph neural network to aggregate information between items into session representations. GC-SAN \cite{xu2019graph} employs a self-attention mechanism to capture item dependencies and user interests via graph information aggregation. GCE-GNN \cite{wang2020global} proposes to capture both global-level and session-level interactions and aggregates item information through graph convolution and self-attention mechanisms. Xia \textit{et al.} \cite{xia2021aaai, xia2021cikm} proposed integrating self-supervised learning into session-based recommendation to boost recommendation performance. Transformer-based session-based recommendation models also achieve promising performance by integrating techniques from Transformer into session-based scenarios \cite{de2021transformers4rec, chen2019bert4sessrec}. A representative method is BERT4SessRec \cite{chen2019bert4sessrec}, which employs BERT for session-based recommendation by using bidirectional encoder representations from Transformer to capture bidirectional correlations in each session effectively. These models are of great capacity in generating accurate recommendations but they were designed for server-side use, which cannot run on resource-constrained devices like smartphones.

\subsection{Knowledge Distillation for Recommendation}
Knowledge Distillation (KD) \cite{hinton2015distilling} is a widely used approach for transferring knowledge from a well-trained large model (teacher) to a simple and lightweight model (student). In this framework, the student model is usually optimized towards two objectives: minimizing the difference between the prediction and the ground truth and fitting the teacher's label distribution or intermediate layer embeddings. KD was first introduced to the classification problem, and currently, some KD methods have been proposed for recommender systems \cite{tang2018ranking, lee2019collaborative, kang2020rrd}. The first work is Ranking Distillation \cite{tang2018ranking}, which chooses top-K items from the teacher's recommendation list as the external knowledge to guide the student model to assign higher scores when generating recommendations. However, this ignores rich information beyond the recommended top-K items. The follow-up work, CD \cite{lee2019collaborative} introduces a rank-aware sampling method to sample items from the teacher model as knowledge for the student model, enlarging the range of information to be transferred. 
DE-RRD \cite{kang2020rrd} develops an expert selection strategy and relaxes the ranking-based sampling method to transfer knowledge. Though effective, these KD methods can hardly tackle the cold-start problem caused by the long-tail distribution in recommender systems. 

\subsection{Compositional Code Learning}
One-hot encoding is a standard technique in deep representation learning, in which each entity is associated with a continuous embedding vector and usually leads to a large embedding table to be stored. To reduce the memory footprint, some research has explored efficient coding systems, for example, Huffman Code \cite{huffman1952method, han2015deep} and Hash functions \cite{tito2017hash, chen2015compressing}. 
However, these coding systems often suffer from low accuracy due to limited representation capability of binary codes. Subsequent works \cite{chen2018learning, shu2017compressing} then explore the use of addictive quantization for source coding, resulting in compositional codes. Specifically, Shu \textit{et al.} \cite{shu2017compressing} proposed an end-to-end compositional encoding-based neural network to compress the word embeddings in NLP, achieving success in compression ratios and in the meantime being independent of languages. Chen \textit{et al.} \cite{chen2018learning} introduced a K-way d-dimensional discrete code for compact embedding representations, which can be generally applied to any differentiable computational graph with an embedding layer. Compositional code has been also explored in recommender systems \cite{liu2019compositional, shi2020compositional, kang2020learning}. {Kang \textit{et al.} \cite{kang2020learning} proposed dense hash encoding to emb categorical features by using multiple hash functions and transformations, which is very close to compositional encoding.}  Shi \textit{et al.} \cite{shi2020compositional} proposed to reduce embedding size by exploiting complementary partitions of category set to produce a unique embedding for each category at the smallest cost. Liu \textit{et al.} \cite{liu2019compositional} represented items/users with a set of binary vectors associated with a sparse weight vector, and an integer weight approximation scheme is proposed to accelerate the speed of the method. But their methods either require category or feature information or bear a high computational cost. Li \textit{et al.} \cite{li2021lightweight} proposed to use compositional code to compress item embedding table in sequential recommendation. However, the model adopts quotient-remainder trick to make the codes of items distinct which increases the time complexity. Besides, its model compression ratio is rather limited. 

\begin{figure}[t]
	\centering
	\includegraphics[width=\textwidth]{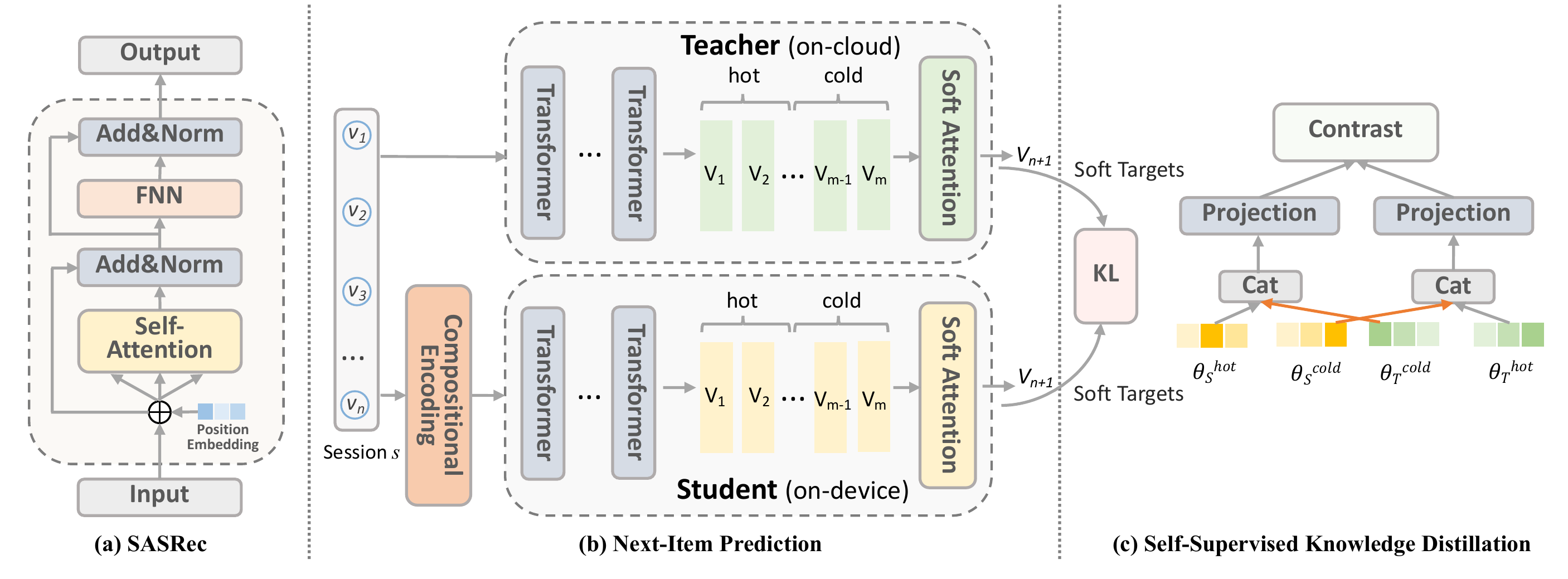}
	\caption{An overview of the proposed method.}
	\label{figure.1}
	\vspace{-10pt}
\end{figure}

\section{Preliminaries}
In this section, we first introduce the session-based recommendation task and then give a brief review of the backbone model and the compression method used in our previous work to help understand the new contributions. 
\subsection{Session-Based Recommendation Task}
Session-based recommender systems rely on the previous clicks in a short period of time to generate the next-item recommendation. Let $\mathcal{V} = \{v_{1}, v_{2}, v_{3}, ... , v_{|\mathcal{V}|}\}$ denote item set and $s = [v_{s,1}, v_{s,2}, ..., v_{s,l}]$ denote a session sequence. Every session sequence is composed of interacted items in the chronological order from an anonymous user. The task of session-based recommendation is to predict the next item, namely $v_{s,l+1}$. In this scenario, every item $v\in \mathcal{V}$ is first mapped into the embedding space. The item embedding table is with a huge size and is denoted as $\mathbf{X}\in\mathbb{R}^{|\mathcal{V}| \times N}$. Given $\mathcal{V}$ and $s$, the output of session-based recommendation model is a ranked list $y = [y_{1}, y_{2}, y_{3}, ..., y_{|\mathcal{V}|}]$ where $y_{v}\ (1 \leq v \leq |\mathcal{V}|)$ is the corresponding predicted probability of item $v$. 
The top-\textit{K} items $(1 \leq K \leq |\mathcal{V}|)$ with highest probabilities in $y$ will be selected as the recommendations.

\subsection{A Brief Review of OD-Rec}
In our previous work OD-Rec, we unify the server-side model (teacher) and the on-device model (student) into a knowledge distillation framework. Both the teacher and the student adopt the Transformer-based sequential recommendation model SASRec \cite{kang2018self} as their backbones to generate next-item recommendation. To make the student fit in resource-constrained devices, we compress its embedding table through the tensor-train decomposition \cite{oseledets2011tensor}. Different from other tensor-train decomposition-based models \cite{wang2020next,yin2021tt}, our student model relaxes the constraint of dimensionality consistency in decomposition so as to achieve a higher compression rate. To retain the capacity of the student, we not only employ the traditional soft targets distillation but also devise two self-supervised distillation tasks to transfer the teacher's knowledge to the student. \par 
\subsubsection{Backbone Model Structure}
The backbone used in our framework is a famous Transformer-based \cite{vaswani2017attention} sequential recommendation model in which several self-attention blocks are stacked, including the embedding layer, the self-attention layer and the feed-forward network layers. The embedding layer adds position embeddings to the original item embeddings to indicate the temporal and positional information in a sequence. The self-attention mechanism can model the item correlations. Inside the block, residual connections, the neuron dropout, and the layer normalization are sequentially used. SASRec with a one-layer setting can be formulated as:
\begin{equation}
\mathbf{\hat{X}} = \mathbf{X} + \mathbf{P},\ \mathbf{F} = \mathrm{Attention}(\mathbf{\hat{X}}),\ \mathbf{\Theta} = \mathrm{FNN}(\mathbf{F}),
\end{equation}
where $\mathbf{X}$, $\mathbf{P}$ are the item embeddings and position embeddings, respectively, $\mathbf{F}$ is the session representation learned via self-attention blocks. $\mathrm{FNN}$ represents feed forward network layers and $\mathbf{\Theta}$ aggregates embeddings of all items. In the origin SASRec, the embedding of the last clicked item in $\mathbf{\Theta}$ is chosen as the session representation. However, we think each contained item would contribute information to the session representation for a comprehensive and more accurate understanding of user interests. Instead, we slightly modify the original structure by adopting the soft-attention mechanism \cite{wang2020global} to generate the session representation, which is computed as:
\begin{equation}
	\alpha_{t}=\mathbf{f}^{\top} \sigma\left(\mathbf{W}_{1} \mathbf{x}^{*}_{s}+\mathbf{W}_{2} \mathbf{x}_{t}+\mathbf{c}\right),\,\,\,
	\mathbf{\theta}_{s}=\sum_{t=1}^{l} \alpha_{t} \mathbf{x}_{t},
\end{equation}
where $\mathbf{W}_{1}\in\mathbb{R}^{N \times N}$, $\mathbf{W}_{2} \in\mathbb{R}^{N \times N}$, $\mathbf{c}, \mathbf{f}\in\mathbb{R}^{N}$ are learnable parameters, $\mathbf{x}_{t}$ is the embedding of item $t$ and $\mathbf{x}_{s}^{*}$ is obtained by averaging the embeddings of items within the session $s$, i.e. $\mathbf{x}_{s}^{*} = \frac{1}{l}\sum_{t=1}^{l}\mathbf{x}_{t}$. $\sigma$ is sigmoid function. Session representation $\mathbf{\theta}_{s}$ is represented by aggregating item embeddings while considering their corresponding importance.

\subsubsection{Model Compression in Our Previous work}
To factorize the embedding table of the student model into smaller tensors, we let the total number of items $|V| = \prod_{k=1}^{d} I_{k}$ and the embedding dimension $N = \prod_{k=1}^{d} J_{k}$. For one specific item $v$ indexed by $i$, we map the row index $i$ into $d$-dimensional vectors $\mathbf{i}$ = $\left(i_1, i_2, ..., i_d\right)$ according to \cite{oseledets2011tensor,hrinchuk2019tensorized}, and then get the particular slices of the index in TT-cores to perform matrix multiplication on the slices. Benefiting from this decomposition, the model just needs to store the TT-cores to fulfill an approximation of the entry in the embedding table, i.e., a sequence of small tensor multiplication. The tensor-train decomposition can be formed as:
\begin{equation}
\mathbf{X} = \mathbf{G}_1 \times \mathbf{G}_2 \times \cdots \times \mathbf{G}_d,
\end{equation}
where $\mathbf{X}$ is a d-dimensional embedding table and $\{\mathbf{G}_{k}\}_{k=1}^{d}$ are called TT-cores, $\mathbf{G}_{k}$ is of the size $\mathbb{R}^{R_{k-1} \times I_k \times J_k \times R_{k}}$ and the sequence of $R_{k}$ is called TT-rank where $R_{0} = R_{d} = 1$ and we set $R_{k} = R_{k-1} = R$. Each entry in $\mathbf{X}$ indexed by ($i_{1}, i_{2}, ..., i_{d}$) can be represented in the following TT-format:
\begin{equation}
\begin{aligned} 
x&=\sum_{r_{1}=1}^{R_{1}} \sum_{r_{2}=1}^{R_{2}}... \sum_{r_{d-1}=1}^{R_{d-1}} \mathbf{G}_{1}\left(i_{1}, r_{1}\right) \mathbf{G}_{2}\left(r_{1}, i_{2}, r_{2}\right)... \mathbf{G}_{d}\left(r_{d-1}, i_{d}\right) \\
&=\underbrace{\mathbf{G}_{1}\left[i_{1},:\right]}_{1 \times R_{1}} \underbrace{\mathbf{G}_{2}\left[:, i_{2},:\right]}_{R_{1} \times R_{2}}...\underbrace{\mathbf{G}_{d-1}\left[:, i_{d-1},:\right]}_{R_{d-2} \times R_{d-1}} \underbrace{\mathbf{G}_{d}\left[:, i_{d}\right]}_{R_{d-1} \times 1}.
\end{aligned}
\end{equation}
However, conventional tensor-train decomposition requires strict dimensionality consistency between factors and probably leads to dimension redundancy.
For example, if we decompose the item embedding matrix into two smaller tensors, i.e., $\mathbf{X} = \mathbf{G}_{1}\mathbf{G}_{2}$, to get each item's embedding, the number of columns of $\mathbf{G}_{1}$ should be the same with the number of rows of $\mathbf{G}_{2}$. 
This consistency constraint is too strict for flexible and efficient tensor decomposition, posing an obstacle to further compression of the embedding table. Therefore, inspired by \cite{zhao2021semi}, we propose to integrate semi-tensor product with tensor-train decomposition where the dimension of TT-cores can be arbitrarily adjusted. Let $\mathbf{a} \in \mathbb{R}^{1 \times np}$ denotes a row vector and $\mathbf{b} \in \mathbb{R}^{p}$, then $\mathbf{a}$ can be split into $p$ equal-size blocks as $\mathbf{a}^{1}$, $\mathbf{a}^{2}$, ..., $\mathbf{a}^{p}$, the left semi-tensor product denoted by $\ltimes$ can be defined as:
\begin{equation}
\left\{\begin{array}{l}\mathbf{a} \ltimes \mathbf{b}=\Sigma_{i=1}^{p} \mathbf{a}^{i} \mathbf{b}_{i} \in \mathbb{R}^{1 \times n} \\ \mathbf{b}^{\mathrm{T}} \ltimes \mathbf{a}^{\mathrm{T}}=\Sigma_{i=1}^{p} \mathbf{b}_{i}\left(\mathbf{a}^{i}\right)^{\mathrm{T}} \in \mathbb{R}^{n}.\end{array}\right.
\end{equation}
Then, for two matrices $\mathbf{A} \in \mathbb{R}^{H \times nP}$, $\mathbf{B} \in \mathbb{R}^{P \times Q}$, STP is defined as:
\begin{equation}
\mathbf{C}=\mathbf{A} \ltimes \mathbf{B},
\end{equation}
and $\mathbf{C}$ consists of $H \times Q$ blocks and each block $\mathbf{C}^{hq} \in \mathbb{R}^{1 \times n}$ can be calculated as:
\begin{equation}
\mathbf{C}^{h q}=\mathbf{A}(h,:) \ltimes \mathbf{B}(:, q),
\end{equation}
where $\mathbf{A}(h,:)$ represents $h$-th row in $\mathbf{A}$, $\mathbf{B}(:, q)$ represents $q$-th column in $\mathbf{B}$ and $h$ = 1, 2, $\cdots$, $H$, $q$ = 1, 2, $\cdots$, $Q$. With the above definitions, the conventional tensor product between TT-cores can be replaced with the semi-tensor product as:
\begin{equation}
\mathbf{X} = \hat{\mathbf{G}}_1 \ltimes \hat{\mathbf{G}}_2 \ltimes \cdots \ltimes \hat{\mathbf{G}}_d,
\end{equation}
where $\{\hat{\mathbf{G}}_{k}\}_{k=1}^{d}$ are the core tensors after applying semi-tensor product based tensor-train decomposition (STTD) and $\hat{\mathbf{G}}_{1} \in \mathbb{R}^{I_1 J_1 \times R}$, $\{\hat{\mathbf{G}}_{k}\}_{k=2}^{d-1} \in \mathbb{R}^{ \frac{R}{n} \times \frac{I_{k} J_k}{n} \times R}, \hat{\mathbf{G}}_{d} \in \mathbb{R}^{\frac{R}{n} \times \frac{I_d J_d}{n}}$.
Semi-Tensor Product loosens the strict dimensionality consistency by enabling the factorized tensors to do the product as long as their corresponding dimensions are proportional. The compression rate is:
\begin{equation}
\text { rate }=\frac{\prod_{k=1}^{d} I_{k} J_{k}}{I_1 J_1 R + \sum_{k=1}^{d-1} I_{k} J_{k} \frac{R^2}{n^2} + I_d J_d \frac{R}{n^2}}.
\end{equation}
We can adjust the TT-rank, namely, the values of the hyperparameters $R$ and $n$, and the length $k$ of the tensor chain to flexibly compress the model to any degree. 

\begin{figure}[t]
	\centering
	\includegraphics[width=.9\textwidth]{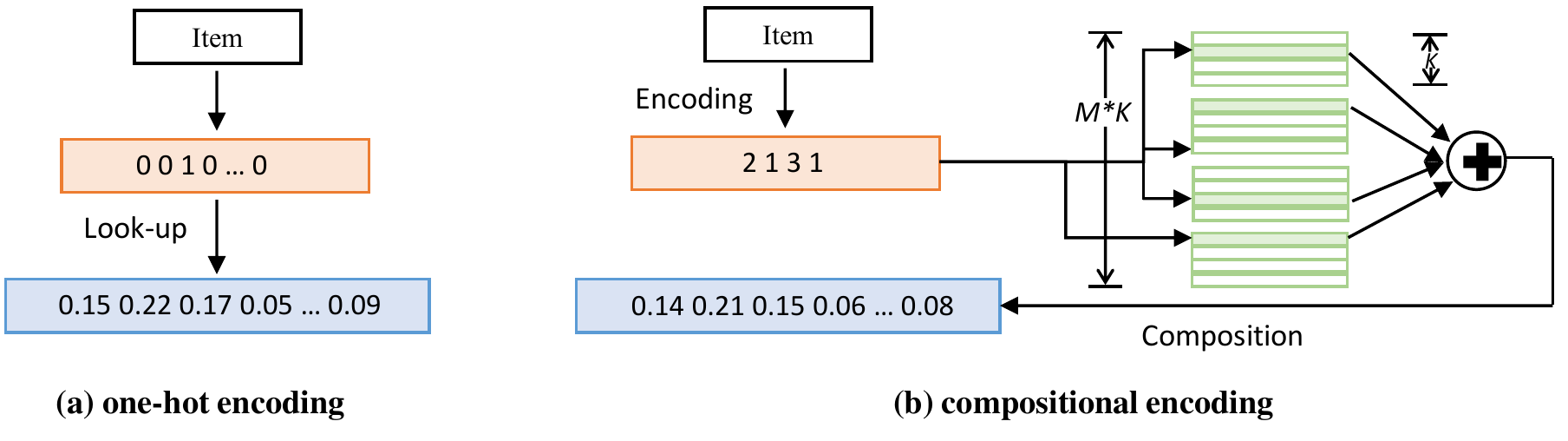}
	\caption{Difference between one-hot encoding and compositional encoding.}
	\label{figure.2}
	\vspace{-10pt}
\end{figure}

\section{Compositional Encoding for Model Compression}
Although the tensor-train decomposition empowered by the semi-tensor product can reduce a substantial memory footprint, the multiplication of tensors also greatly increases the inference time. In this section, we present our new compression method which is based on the compositional code learning. 
\subsection{Discrete Code Embedding Framework}
We propose to compress the item embedding table by enabling items to share embedding vectors through a discrete code learning mechanism. Specifically, each item $v$ will be represented by a unique code $C_v = (C_v^1, C_v^2, ..., C_v^M) \in \mathcal{B}^{M}$ where each component in the code is a discrete number ranged in [1, $K$] and $\mathcal{B}$ is the set of code bits with cardinality of $M$.  There is a code allocation function $\Phi(\cdot): \mathcal{V} \rightarrow \mathcal{B}^{M}$ that maps each item with its discrete code. Since the discrete code has $M$ values, we cannot directly use the code to look up the item vector in the embedding table; instead, we expect to learn a code composition function $\Theta$ that takes a discrete code of item $v$ as input and generates the corresponding continuous embedding vector as output, i.e. $\Theta: \mathcal{B}^{M} \rightarrow \mathbb{R}^{N}
$. Then the discrete code framework can be formed as:
\begin{equation}
\Gamma = \Phi \cdot \Theta.
\end{equation}
Given learned $\Phi$ and item $v$, we can get its code, i.e. $C_v = \Phi (v) = (C_v^1, C_v^2, ..., C_v^M)$. In order to get the composite embedding, we adopt a code composition function $\mathbf{e}_v = \Theta(C_v)$.  We first embed $C_v$ to a sequence of embedding vectors $(\bm{E}_1^{C_v^1}, \bm{E}_2^{C_v^2}, ..., \bm{E}_i^{C_v^i})$. And then we apply embedding transformation to generate $\bm{e}_v$.
Here we create $M$ codebooks $\bm{E}_1, \bm{E}_2, ..., \bm{E}_M$, each containing $K$ vectors with the dimensionality of $N$. The final embedding of item $v$ is 
\begin{equation}
	\bm{e}_v=\sum_{i=1}^{M} \bm{E}_{i}\left(C_{v}^{i}\right),
\end{equation}
where $\bm{E}_{i}\left(C_{v}^{i}\right)$ is the $C_{v}^{i}$-th vector in codebook $\bm{E}_i$. Therefore, in this case, we just need to store the integer codes of all items, which is called code matrix $C$ and the $M$ codebooks. We present the processes of one-hot encoding and compositional encoding in Fig. \ref{figure.2}. For one-hot encoding, the number of vectors used in representing items is identical to the number of items. In contrast, the number of vectors for compositional approach required to construct the embeddings is $M*K$. To store each code, $M \log _2 K$ bits are required so $K$ is selected to be a number of a multiple of 2 for the convenience of storage. \par
In this case, the compression ratio can be calculated as:
\begin{equation}
	\text { rate }=\frac{|\mathcal{V}| \cdot N}{MKN + M\cdot|\mathcal{V}|}=\frac{N}{\frac{MN}{|\mathcal{V}|}K + M}.
\end{equation}
We can adjust the value of $M$ and $K$ to flexibly compress the model to any degree. As shown in the above equation, the upper bound of the compression ratio is $N/M$. Therefore, when $N$ is fixed, we should seek small $M$ to have a larger compression ratio. Here we show the varying model compression ability with different values of $M$ and $K$ in Table \ref{Table.1}. We assume that there are 20,000 items and the dimension of each embedding vector is 100. Then we calculate the model size before and after model compression and present them in Table \ref{Table.1}. It is clear that the value of $M$ affects the compression ratio more than $K$ because $MN\ll |\mathcal{V}|$. 
\begin{table}
	\centering
	\begin{tabular}{c|c|c|c|c} 
	\hline
	\makecell{Original size of\\ item embeddings}  & \makecell{Codebook\\number M} & \makecell{Vector\\number K}   & \makecell{Size after compression} & \makecell{Compression rate}   \\\hline
				  &   & \makecell[c]{8}   &  41600    &    48 \\
				  2,000,000	  & 2 & \makecell[c]{128}  &   65600   &30     \\
				  &   & 512 &    142400  &   14  \\\hline
				  &   & \makecell[c]{8}    &   83200   &24     \\
	2,000,000     & 4 & \makecell[c]{128}  &    131200  & 15    \\
				  &   & 512 &   284800   &  7   \\\hline
				  &   & \makecell[c]{8}    &     185600 & 11    \\
				  2,000,000		  & 8 & \makecell[c]{128}  &    262400  & 8    \\
				  &   & 512 &   569600   & 3    \\
	\hline
	\end{tabular}
	\caption{Model Size Analytics with Different $M$ and $K$.}	
	\label{Table.1}
	\vspace{-20pt}
	\end{table}

\subsection{Compositional Code learning}
Although the embedding size can be greatly reduced after using discrete code, we expect to prevent serious performance degradation and maintain model capacity in next-item recommendation. Namely, given baseline embedding matrix $\mathbf{X}$ (Here the baseline embedding is the well-trained item embedding matrix in the teacher model because we have a knowledge distillation framework), we want to find a set of codes $C$ and codebooks $\bm{E}$ that can generate embeddings with the same effectiveness as $\mathbf{X}$. A straightforward way is to minimize the squared distance between the original embedding and the composite embedding:
\begin{equation}
	\begin{aligned}
	\mathcal{L}_{mse} &=\underset{C, \bm{E}}{\operatorname{argmin}} \frac{1}{|\mathcal{V}|} \sum_{v \in \mathcal{V}}\left\|\bm{E}(C_{v})-\mathbf{X}_v\right\|^{2} \\
	&=\underset{C, \bm{E}}{\operatorname{argmin}} \frac{1}{|\mathcal{V}|} \sum_{v \in \mathcal{V}}\left\|\sum_{i=1}^{M} \bm{E}_{i}(C_{v}^{i})-\mathbf{X}_v\right\|^{2}.
	\end{aligned}
\end{equation}

\subsubsection{End-To-End Learning with Gumbel-Softmax}
To minimize the MSE loss means we should optimize the item-to-code mapping function $\Phi$ and the code-to-embedding composition function $\Theta$. However, each code is discrete so the learning is not differentiable. In order to enable an end-to-end learning, we consider that each code $C_v$ can be seen as a concatenation of $M$ one-hot vectors, i.e. $C_v = (\bm{O}_v^1, \bm{O}_v^2, ..., \bm{O}_v^M)$, where $\forall i, \bm{O}_{v}^{i} \in[0,1]^{K}$ and $\sum_{k} \bm{O}_{v}^{i k}=1$, and $\bm{O}_{v}^{i k}$ is the $k$-th component of $\bm{O}_v^i$. Let $\bm{O}^1, \bm{O}^2, .., \bm{O}^M$ represent $M$ code matrices where each matrix is with the size of $|V| \times K$ and each row in each matrix is a $K$-dimensional one-hot vector. 
Then the generation of item's embedding can be reformulated to be:
\begin{equation}
	\bm{e}_v=\sum_{i=0}^{M} \bm{E}_{i}^{\top} \bm{O}_{v}^{i},
\end{equation}
where $\bm{O}_v^i$ represents the one-hot vector corresponding to the code component $C_v^i$ of item $v$. Therefore, the optimization of the item-to-code-embedding function becomes to find an optimal set of one-hot code matrices $\bm{O}^1, \bm{O}^2, .., \bm{O}^M$ and basis codebooks $\bm{E}_1, \bm{E}_2, .., \bm{E}_M$, that minimize the reconstruction loss $\mathcal{L}_{mse}$ mentioned above. 
Inspired by \cite{shu2017compressing}, we then adopt Gumbel-Softmax \cite{JangGP17} to make the continuous vector $\bm{O}_v^i$ approximate the one-hot vector. The $k$-th element in $\bm{O}_v^i$ is computed as:
\begin{equation}
	\begin{aligned}
	\left(\bm{O}_{v}^{i}\right)_{k} &=\operatorname{softmax}\left(\log \boldsymbol{\alpha}_{v}^{i}+G\right)_{k} \\
	&=\frac{\exp \left(\left(\log \left(\boldsymbol{\alpha}_{v}^{i}\right)_{k}+G_{k}\right) / \varepsilon\right)}{\sum_{k^{\prime}=1}^{K} \exp \left(\left(\log \left(\boldsymbol{\alpha}_{v}^{i}\right)_{k^{\prime}}+G_{k^{\prime}}\right) / \varepsilon \right)},
	\end{aligned}
\end{equation}
where $G_k$ is a noise term that is sampled from the Gumbel distribution $-\log (-\log ($Uniform$(0,1))$, $\varepsilon$ is the temperature in softmax. As $\epsilon\rightarrow$ 0, the softmax computation smoothly approaches the $\arg$max. (In our setting, we empirically set $\varepsilon$ to 0.3, which is reported as a good choice in many previous studies). And $\boldsymbol{\alpha}_{v}^{i}$ is computed by a two-layer MLP:  
\begin{equation}
	\begin{aligned}
	\boldsymbol{\alpha}_{v}^{i} &=\operatorname{softmax}\left(\operatorname{softplus}(\boldsymbol{\theta}_{i}^{\prime \top} \boldsymbol{h}_{v}+\boldsymbol{b}_{i}^{\prime})\right) \\
	\boldsymbol{h}_{v} &=\tanh \left(\boldsymbol{\theta}^{\top} \mathbf{X}_{v}+\boldsymbol{b}\right),
	\end{aligned}
\end{equation}
where $\mathbf{X}_{v}$ is the well-trained embedding of item $v$, $\boldsymbol{\theta} \in  \mathbb{R}^{N \times (MK/2)}$, $\boldsymbol{\theta}_{i}^{\prime} \in \mathbb{R}^{(MK/2) \times MK}$, $\boldsymbol{b} \in \mathbb{R}^{MK/2}$ and $\boldsymbol{b} \in \mathbb{R}^{MK}$. By using this reparameterization trick, the model can circumvent the indifferentiable look-up operation and allows the gradients from the MSE loss to be delivered to the codebooks.

\subsubsection{Code Learning with Guidances}
Although compositional encoding can compress the embedding table, it will definitely degrade recommendation performance. 
In order to retain the model capacity and guide the compositional code learning, we propose \textit{embedding mixup} in the process of code learning. Specifically, during training, instead of solely using the generated embedding from Eq. (15), we use an interpolation of the composite embedding $\bm{e}_v$ and the uncompressed embedding $\mathbf{X}_{v}$:
\begin{equation}
	\bm{e}_v^{\prime}= \eta \odot \mathbf{X}_{v}+(1-\eta) \odot \bm{e}_v,
\end{equation}
where $\eta \in (0,1)$ is a hyperparameter. We can tune it to achieve the best performance. Note that we only use the embedding mixup in training. In the test/inference phase, only the compositional embedding is used for prediction.

\subsubsection{Time Complexity Analysis}
In this section, we compare the time complexity of the proposed new compression method and the previous tensor-train decomposition. The discussion focuses on the process of reconstructing item embeddings, which is critical for fast local inference. Recall that, in tensor-train decomposition, for each item there is an index list to calculate, followed by a series of tensor multiplications between the vectors looked up from factorized tensors with the index list. In our new method, to generate item embeddings, we just need the matrix multiplication between the code matrix and codebooks. Here we let $|V|$ be the number of items, $N$ is the embedding dimension, $R$ is value of TT-rank in tensor-train decomposition, $M$ is the number of codebooks and $K$ is the number of vectors in each codebook. In tensor-train decomposition, the time complexity of the index list calculation for $n$ items is $\mathcal{O}(|V|^2 + |V|)$; as for the multiplication between indexed embedding vectors, its time complexity is $\mathcal{O}(|V|R)$. Then the total time complexity of tensor-train decomposition is $\mathcal{O}(|V|^2 + |V| + |V|R)$. When using compositional codes to represent items, we first pre-train the code matrix and codebooks and then they are saved to form item embeddings. Therefore, there is only one-time matrix multiplication between the code matrix and the codebooks, and the time complexity is $\mathcal{O}(|V|MKN)$. Since $M$ and $K$ are small, the new compression method is theoretically much more efficient than the old one.

\section{Bidirectional Self-Supervised Knowledge Distillation}
In our previous paper, we proposed a self-supervised knowledge distillation framework to improve the capacity of the on-device model in a teacher-student fashion. In this paper, we advance it by proposing bidirectional knowledge distillation. 
\subsection{Framework Formulation}
Since we are the first to combine self-supervised learning and knowledge distillation for recommendation, we define the general framework of self-supervised knowledge distillation for recommendation as follows. Let $\mathcal{D}$ denote the item interaction data, $\mathcal{D}_{\mathrm{soft}}$ represent the soft targets from the teacher and $\mathcal{D}_{\mathrm{aug}}$ denote the augmented data under the self-supervised setting. The teacher model and the student model are denoted by $M_{T}$ and $M_{S}$, respectively. Then the framework is formulated as follows:
\begin{equation}
f\left(M_{T}(\mathcal{D}), M_{S}(\mathcal{D}, \mathcal{D}_{\mathrm{soft}}, \mathcal{D}_{\mathrm{aug}}) \right) \rightarrow M_{S}^{*},
\end{equation}
which means that by jointly supervising the student model with the historic interactions $\mathcal{D}$, the teacher's soft targets $\mathcal{D}_{\mathrm{soft}}$ and the self-supervised signals extracted from $\mathcal{D}_{\mathrm{aug}}$, we can finally obtain an improved student model $M_{S}^{*}$ that retains the capacity of the regular model. 
\par
\subsection{Data Augmentation}
In this framework, two types of distillation tasks: traditional distillation task based on soft targets, and contrastive self-supervised distillation task are integrated.
The essential idea of self-supervised learning \cite{liu2020self,yu2021socially,yu2021www} is to learn with the supervisory signals which are extracted from augmentations of the raw data. However, in recommender systems, interactions follow a long-tail distribution. For the long-tail items with few interactions, generating informative augmentations is often difficult \cite{yu2022survey}. 
Inspired by the genetic recombination \cite{meselson1975general} and the preference editing in \cite{ma2021improving}, we come up with the idea to exchange embedding segments from the student and the teacher and recombine them to distill knowledge. Such recombinations enable the direct information transfer between the teacher and the student, and meanwhile create representation-level augmentations which inherit characteristics from both sides. Under the direct guidance from the teacher, we expect that the student model can learn more from distillation tasks. We term this method \textit{embedding recombination}. 
\par
To implement it, we divide items into two types: hot and cold by their popularity. The top 20\% popular items are considered as hot items and the rest part is cold items. For each session, we split it into two sub-sessions: cold session and hot session in which only the cold items or hot items are contained. Then we learn two corresponding session representations respectively: hot session representation and cold session representation. The hot session representation derives from representations of hot items in that session based on the soft attention mechanism in Eq. (3), while the cold session representation is analogously learned from representations of cold items. They are formulated as follows:
\begin{equation}
\mathbf{\theta}_{s}^{hot}=\sum_{t=1}^{l_{h}} \alpha_{t}^{hot} \mathbf{x}_{t}^{*hot}, \,\,\,\mathbf{\theta}_{s}^{cold}=\sum_{t=1}^{l_{c}} \alpha_{t}^{cold} \mathbf{x}_{t}^{*cold},
\end{equation}
where $l_{c}$ and $l_{h}$ are lengths of the cold session and the hot session, $\mathbf{x}_{t}^{*hot}(\mathbf{x}_{t}^{*cold})$ represents representation of the t-$th$ hot item or cold item in the session $s$, $\alpha_{t}^{hot}(\alpha_{t}^{cold})$ is the learned attention coefficient of the t-$th$ hot or cold item in the session, and $\mathbf{\theta}^{hot}(\mathbf{\theta}^{cold})$ is the learned hot or cold session representation. In both the student and the teacher models, we learn such session embeddings. Then for the same session, we swap their cold session representations as follows to generate new embeddings:
\begin{equation}
z_{s}^{tea} =[\mathbf{\theta}_{s}^{hot_{tea}}, \mathbf{\theta}_{s}^{cold_{stu}}],\,\,\, z_{s}^{stu} = [\mathbf{\theta}_{s}^{hot_{stu}}, \mathbf{\theta}_{s}^{cold_{tea}}].
\end{equation}
As for those sessions which only contain hot items or cold items, we generate the corresponding type of session representations and swap them.

\subsection{Knowledge Distillation Tasks}
\subsubsection{Contrastive Distillation.}
Since the generated new embeddings are profiling the same session, we consider that there is shared learnable invariance lying in these embeddings. Therefore, we propose to contrast recombined embeddings so as to learn discriminative representations for the student model, which can mitigate the data sparsity issue to some degree. Meanwhile, this contrastive task also helps to transfer the information in the teacher's representation, rather than only transferring the output probabilities. We follow \cite{yu2022graph,yu2022xsimgcl} to use InfoNCE \cite{oord2018representation} to maximize the mutual information between teacher's and student's session representations. For any session $s$, its recombined representations are positive samples of each other, while the recombined representations of other sessions are its negative samples. We conduct negative sampling in the current batch $\mathcal{B}$. The loss of the contrastive task is defined as follows:
\begin{equation}
\mathcal{L}_{cl}=-\underset{s\in\mathcal{B}}{\sum}\log \frac{ \psi\left(\mathbf{W}_{t} z_{s}^{tea}, \mathbf{W}_{s} z_{s}^{stu}\right)}{ \psi\left(\mathbf{W}_{t} z_{s}^{tea}, \mathbf{W}_{s} z_{s}^{stu}\right)+\underset{j\in\mathcal{B}/\{s\} }{\sum} \psi\left(\mathbf{W}_{t} z_{s}^{tea}, \mathbf{W}_{s} z_{j}^{stu}\right)},
\end{equation}
where $\psi\left(\mathbf{W}_{t} z_{s}^{tea}, \mathbf{W}_{s} z_{s}^{stu}\right)$ = $\exp\left(\phi\left(\mathbf{W}_{t} z_{s}^{tea}, \mathbf{W}_{s} z_{s}^{stu}\right) / \tau\right)$, $\phi\left(\cdot\right)$ is the cosine function, $\tau$ is temperature (0.2 in our method), and $\mathbf{W_{t}}$ and $\mathbf{W_{s}}\in \mathbb{R}^{N \times 2N}$ are projection matrices.  \par

\subsubsection{Soft Targets Distillation.}The self-supervised KD tasks rely on data augmentations. Though effective, there may be divergence between original representations and recombined representations. We finally distill the soft targets (i.e., teacher's prediction probabilities on items). Following the convention, we adopt KL Divergence to make the student generate probabilities similar to the teacher's. Let $\mathrm{prob}^{tea}$ and $\mathrm{prob}^{stu}$ represent the predicted probability distributions of the teacher and the student, respectively. Then for each session $s$, we have:
\begin{equation}
\begin{aligned}
\mathrm{prob}^{tea} = \mathrm{softmax}(\mathbf{\theta}_{s}^{tea\top}\mathbf{X}),\ \ \ 
\mathrm{prob}^{stu} = \mathrm{softmax}(\mathbf{\theta}_{s}^{stu\top}\mathbf{X});
\end{aligned}
\end{equation}
$\mathbf{\theta}_{s}^{tea}$ and $\mathbf{\theta}_{s}^{stu}$ are representations of session $s$ from teacher and student. Then to maximize the agreement between them, we define the loss of distilling soft targets as:
\begin{equation}
\mathcal{L}_{\mathrm{soft}} = - \sum_{v=1}^{|\mathcal{V}|} \mathrm{prob}^{tea}_v \ln \frac{\mathrm{prob}^{stu}_v}{\mathrm{prob}^{tea}_v}.
\end{equation}
\subsection{Bidirectional Knowledge Distillation}
The self-supervised knowledge distillation framework has been demonstrated to be effective in our previous work. However, the framework also has limitations. To be specific, the framework is based on the assumption that the teacher is superior than the student so the teacher can transfer its knowledge to help the student. But in practice the teacher is not always superior than the student and sometimes the student can even achieve better performance, which has been reported in the results in our previous work. Such a phenomenon is also observed in \cite{kweon2021bidirectional}. The possible reason is that the teacher may mistakenly rank some items which should be recommended low whereas the student can correctly rank them high. Therefore, in this work we conduct bidirectional knowledge distillation which allows the reverse transfer knowledge from the student to the teacher. In this way, the two models can recursively learn from each other and get improved. Since both the teacher and the student are trainable, the upgraded framework is reformulated as:
\begin{equation}
	f\left(M_{T}(\mathcal{D}, \mathcal{D}_{\mathrm{soft}}, \mathcal{D}_{\mathrm{aug}}), M_{S}(\mathcal{D}, \mathcal{D}_{\mathrm{soft}}, \mathcal{D}_{\mathrm{aug}}) \right) \rightarrow M_{S}^{*}, M_{T}^{*}.
\end{equation}

\subsection{Training Scheme and Model Optimization}
We first train the teacher model in the cloud where the embedding table is not compressed. Then, the student model is trained under the help of the teacher in the cloud. Finally, the well-trained lightweight student model is downloaded and saved on the device for local inference. Before training with the knowledge distillation framework, we first pre-train the student with the recommendation task to get the well-trained code matrix and codebooks. The pre-training objective can be formed as:
\begin{equation}
	\mathcal{L}_{pre} = \mathcal{L}_{rec} +  \mathcal{L}_{mse}.
\end{equation}
As for the recommendation task, we use the inner product of the candidate item embeddings $\mathbf{E}^{'}$ (obtained through embedding composition in Eq. (10) and embedding mixup in Eq. (16)) and the given session $s$ to denote the scores of items, then we apply softmax to compute the probabilities of items to be recommended:
\begin{equation}
	\hat{\mathbf{y}}=\operatorname{softmax}(\mathbf{\theta}_{s}^{\top}\bm{E}^{'}),
\end{equation}
where $\mathbf{\theta}_{s}$ is computed through Eq. (3).
The cross-entropy is used as the loss function of the recommendation task:
\begin{equation}
	\mathcal{L}_{rec}=-\sum_{v=1}^{|V|} \mathbf{y}_{v} \log \left(\hat{\mathbf{y}}_{v}\right)+\left(1-\mathbf{y}_{v}\right) \log \left(1-\hat{\mathbf{y}}_{v}\right)
\end{equation}
where $\mathbf{y}$ is the one-hot encoding vector of the ground truth. For simplicity, we leave out the $L_{2}$ regularization terms. After this warm-up, we unify the teacher and the student into the bidirectional self-supervised knowledge distillation framework. 
Then the total loss is:
\begin{equation}
	\mathcal{L} = \mathcal{L}_{rec} +  \mathcal{L}_{mse} + \mathcal{L}_{kd}
	= \mathcal{L}_{rec} +  \mathcal{L}_{mse} + \beta \mathcal{L}_{con} + \gamma \mathcal{L}_{\mathrm{soft}},
\end{equation}
where $\beta$, $\gamma$ are coefficients that control the magnitudes of two knowledge distillation tasks. The whole training process is presented in Algorithm 1.

\begin{algorithm}[t]
    \caption{The training process of the proposed method.}
    \LinesNumbered 
    \label{alg:Framework}
    
    \KwIn{session data $\mathcal{D}$}
    \KwOut{Recommendation lists} 
	Train the teacher model $M_T$ on the cloud and get the well-trained item embedding matrix $\mathbf{X}$\;     
    Pre-train the student with Eq. (24) to get the codebooks $\bm{E}_1, \bm{E}_2,...\bm{E}_M$ and code matrix $C$ for items\;
    \For {each iteration}{
        Form item embedding through Eq. (13)-Eq. (16)\;
		Compute $\mathcal{L}_{mse}$ through Eq. (12)\;
        Form session representations through Eq. (18) - Eq. (19)\;
		Compute the recommendation loss by following Eq. (25) and (26)\;
		Compute the knowledge distillation loss by following Eq. (20) - Eq. (22)\;
		Jointly optimize the model using Eq. (27) to update both $M_T$ and $M_S$.
        }
\end{algorithm}  

\begin{table}[ht]
		\renewcommand\arraystretch{1.0}
		
		\begin{center}
			\begin{tabular}{c|c|c|c|c}
				\hline
				Dataset&\#Training sessions & \#test sessions &  \#Items & Avg. Length\\ \hline
				\hline			
				Tmall & 351,268 & 25,898 & 40,728 & 6.69\\
				RetailRocket&433,643  &15,132 & 36,968 & 5.43\\
				\hline
			\end{tabular}
		\end{center}
		\caption{Dataset Statistics}	
		\label{Table.2}
		\vspace{-20pt}
	\end{table}

\section{Experiments}
In this section, we first describe the experimental settings, including datasets, baselines, and hyperparameter settings in Section 6.1. Then we evaluate recommendation performances of all baseline methods in Section 6.2. In Section 6.3, we explore the effectiveness of the compressed model by using different values of key hyperparameters. To investigate the on-device inference efficiency of the proposed model, we compare it with other baselines methods in Section 6.4. We also design several variants of the proposed method to verify the contributions of each part in the model in Section 6.5 and investigate the sensitivity of three hyperparameters in the model in Section 6.6. Finally, we analyse the distribution of compositional codes to evaluate if a good code matrix is learned.
\subsection{Experimental Settings}
\subsubsection{Datasets.}
We evaluate our model on two real-world benchmark datasets: \textit{Tmall}\footnote{https://tianchi.aliyun.com/dataset/dataDetail?dataId=42} and \textit{RetailRocket}\footnote{https://www.kaggle.com/retailrocket/ecommerce-dataset}.
\textit{Tmall} is from IJCAI-15 competition and contains anonymized users' shopping logs on Tmall shopping platform. \textit{RetailRocket} is
a dataset on a Kaggle contest published by an E-commerce company, including the user's browsing activities within six months. 
For convenient comparison, we duplicate the experimental environment in \cite{wu2019session, wang2020global}. Specifically, we filter out all sessions whose length is 1 and items appearing less than 5 times. The latest interacted item of each session is assigned to the test set and the previous data is used for training. The validation set is randomly sampled from the training set and makes up 10\% of the training set. Then, we augment and label the training and test datasets by using a sequence splitting method, which generates multiple labeled sequences with the corresponding labels $([v_{s,1}], v_{s,2}), ([v_{s,1},v_{s,2}], v_{s,3}), ...,
([v_{s,1}, v_{s,2}, ..., v_{s,l-1}], v_{s,l})$ for every session $s = [v_{s,1}, v_{s,2}, v_{s,3}, ..., v_{s,l}]$. Note that the label of each sequence is the last consumed item in it. The statistics of used datasets are presented in Table \ref{Table.2}.

\subsubsection{Baseline Methods.}
We compare our method with the following representative session-based recommendation methods (since our work is the first on-device session-based recommendation model, we only compare the new method with its predecessor OD-Rec at the on-device level):
\begin{itemize}[leftmargin=*]
	\item \textbf{GRU4Rec} \cite{hidasi2015session} is a GRU-based session-based recommendation model which also utilizes a session-parallel mini-batch training process and adopts ranking-based loss functions to model user sequences.
	\item \textbf{NARM} \cite{li2017neural} is an RNN-based model which employs an attention mechanism to capture users' main purpose and combines it with the temporal information to generate recommendations.
	\item \textbf{STAMP} \cite{liu2018stamp} adopts attention layers to replace all RNN encoders and employs the self-attention mechanism \cite{vaswani2017attention} to model long-term and short-term user interests.
	\item \textbf{SR-GNN} \cite{wu2019session} proposes a gated graph neural network to refine item embeddings and also employs a soft-attention mechanism to compute the session embeddings.
        \item {\textbf{SASRec} \cite{kang2018self} is Transformer-based sequential recommendation model to capture long-term semantics and predict based on relatively few actions.}
	\item \textbf{OD-Rec} \cite{xia2022device} is our previous work that proposes a session-based recommendation model operating on resource-constrained devices where a tensor-train decomposition method is used to compress the deep model and a self-supervised knowledge distillation framework is proposed to retain model capacity.
\end{itemize}

We use P@K (Precision) and NDCG@K (Normalized Discounted Cumulative Gain) to evaluate the recommendation results where K is 5 or 10. Precision measures the ratio of hit items and NDCG assesses the ranking quality of the recommendation list.

\subsubsection{Hyperparameter Settings}
As for the setting of the general hyperparameters, we set the mini-batch size to 100, the $L_2$ regularization to $10^{-5}$, and the embedding dimension to 128 for Tmall and 256 for RetailRocket. All the learnable parameters are initialized with the Uniform Distribution $U(-0.1,0.1)$. For the backbone, the best performance is achieved when the number of attention layers and attention heads are 1 and 2 for RetailRocket and 1 and 1 for Tmall. The dropout rate is 0.5 for Tmall and 0.2 for RetialRocket. We use Adam with  learning rate 0.001 to optimize the model. We empirically set the temperature in Gumbel-Softmax to 0.3. For all baselines, we report their best performances with the same general experimental settings.

\begin{table}[t]
	\begin{center}
		\setlength{\tabcolsep}{1mm}{
		{
		{
			\begin{tabular}{*{9}{c}}
				\toprule
				\multirow{2}{*}{Method} &
				\multicolumn{4}{c}{Tmall} & \multicolumn{4}{c}{RetailRocket} \cr
				\cmidrule(lr){2-5}\cmidrule(lr){6-9} & Prec@5 & NDCG@5 & Prec@10 & NDCG@10 & Prec@5 & NDCG@5 & Prec@10 & NDCG@10  \\ \hline		
							
				GRU4REC  & 16.48 & 9.25 & 19.58 & 9.93 & 32.37 &  19.36& 39.35 &20.53  \\
				
				NARM & 16.84 & 10.96 & 19.68 &11.66 & 32.94 & 20.07  & 39.18 & 21.53   \\
				
				STAMP  & 17.45 & 12.92  &22.61  &13.34  & 33.57 & 20.50 & 39.75 &  22.97 \\
				
				SR-GNN & 19.39 & 14.29 & 23.79 & 15.81 & 35.68 & 27.19 & 43.31 & 29.67  \\
                    SASRec & 20.23&15.11 &25.13 &16.59 & 36.37& 27.54& 43.55& 29.20\\
				Teacher & {22.47} & {18.05} & {27.86} & {19.53} &{36.51} &{27.77}&{43.59}&{30.03}\\ \hline

				OD-Rec & {23.68} & {18.86} & {25.22} & {19.43} &{35.17} &{28.45}&{36.81}&{28.93}\\ 
				Student & {23.66} & {18.49} & {27.76} & {19.67} &{37.72} &{28.60}&{44.55}&{30.83}\\ \hline			
		\end{tabular}}}}		
	\end{center}
	\caption{Performance comparison for all the methods.}
	\label{Table.3}
	\vspace{-20pt}	
\end{table}
\subsection{Model Performance}
We first report the performances of all methods in Table \ref{Table.3}. We use Teacher to denote the separately trained server-side model and use Student to denote the on-device model trained with the bidirectional self-supervised knowledge distillation framework. OD-Rec is the predecessor of the student model proposed in our previous work. The performance of on-devices model varies drastically with different hyperparameters. We present the results of OD-Rec and the student when they are compressed to the same degree (compression ratio is 30). According to the table, we can draw following conclusions: 
\begin{itemize}[leftmargin=*]
		\item Transformer-based models (SASRec, the teacher model and two student models) outperform all other deep models (RNN-based methods and graph-based methods), showing Transformer-based architecture's superiority in modeling session-based data. Besides, the teacher model and two student models have higher performances than SASRec, verifying that soft attention mechanism we use in the base model helps predict accurate recommendation lists. 
		\item OD-Rec can outperform the teacher on Prec@5 on Tmall and NDCG@5 on two datasets with 30x smaller size. The new student model is further strengthened that it can even outperform the teacher on almost all the metrics and datasets. It also significantly beats OD-Rec especially on Prec@10 on Retailrocket, which demonstrates the effectiveness of the proposed compositional encoding and bidirectional self-supervised knowledge distillation framework.		
	\end{itemize}

\begin{table}[t]
		\footnotesize
	\begin{center}
		\setlength{\tabcolsep}{1mm}{
		{
		{
			
			\begin{tabular}{ccccccc|ccccccc}
				\toprule
				\multicolumn{7}{c}{\textbf{Tmall}} & \multicolumn{7}{c}{\textbf{RetailRocket}} \\\hline
				M & K & CR & Prec@5 & NDCG@5 & Prec@10 & NDCG@10& M & K & CR & Prec@5 & NDCG@5 & Prec@10 & NDCG@10 \\ \hline	
				1 & 32 & 116 &21.10 &15.58 &25.48 & 17.01   &1 &32 & 210&33.09  & 24.41& 40.01&26.65   \\
				1 & 128 & 91 &22.01 &15.48 &26.22 &17.87    &1 & 128&135 & 34.98 &24.60 &41.67 & 27.76 \\
				1 & 256 &71 &22.77 &17.13 &27.02 &18.16     &1& 256&92 & 35.09 &26.46 &42.74 & 27.61 \\
				1 & 512 &49 &23.10 &17.76 &27.38 &18.22     &1 &512 & 56& 36.11& 26.39&42.28 &28.72     \\
				\hline
				2 & 32 &58 &23.01 &17.58 &27.37 & 19.00     &2 &32 &105 & 34.09 & 24.47&41.14 &27.76     \\
				2 & 128 &46 &23.18 &17.63 &27.57 &19.05     &2 & 128& 68& 36.04 & 26.60& 41.76&28.78    \\
				2 & 256 & 35&23.21 &17.80 &27.53 & 19.12    &2& 256&46 &36.10  &26.41 &41.32 &28.65   \\
				2 & 512 &24 &23.45 & 18.65& 27.92& 19.79    &2 &512 & 28& 36.86 &27.40 &43.45 &30.66     \\
				\hline
				3 & 32& 39&23.05 & 17.67& 27.36&19.07       &3 &32 & 70& 36.08 & 26.62& 41.70&27.73     \\
				3 &128 &30 &23.44 &18.43 & 27.42& 19.66     & 3&128 &45 &37.08  & 27.72&43.47 &30.79    \\
				3 & 256& 24&23.75 &18.63 &27.66 & 19.75     &\textbf{3} & \textbf{256}&\textbf{31} &\textbf{37.72}  & \textbf{28.60}& \textbf{44.55}&\textbf{30.83}     \\
				3 & 512& 16&23.82 & 18.89&28.00 &19.84      & 3& 512& 19&37.65  &28.98 &44.37 &30.96      \\
				\hline
				\textbf{4} & \textbf{32} & \textbf{29} &\textbf{23.66} &\textbf{18.49} &\textbf{27.76} &\textbf{19.67} & 4& 32&52 &36.19  &26.67 &42.14 & 28.69      \\
				4 & 128&23 &23.88 &18.45 &27.33 &19.30      &4 &128 &34&37.02  &28.37 &44.25 &30.70     \\
				4 &256 & 18& 23.82& 18.50& 27.86&19.98      & 4&256 &23 &37.53  &28.44 &44.30 &30.79     \\
				4 & 512&12 &23.89 &18.61 &28.04 &19.64      &4 & 512& 14&37.99  &28.83 &44.75 &31.01      \\
				\hline
				5 & 32& 23 &23.57 &18.42 &27.49  & 19.07    &5 &32 &42 & 36.75 &28.51 &42.74 &30.45     \\
				5 & 128& 18 &23.67 &18.51 &27.41 & 18.95     &5 &128 &27 & 37.16 &28.38 &44.23 & 30.67    \\
				5 & 256&  14& 23.88&18.67 &27.44 & 19.06    &5 &256 &18& 37.77 &  28.88& 44.76&30.99     \\
				5 & 512&10  & 23.81&18.89 & 28.13& 19.70     &5 & 512& 11& 37.99  & 28.97& 44.66& 30.78 \\
				\midrule			
		\end{tabular}}}}		
	\end{center}
	\caption{Performance comparison between different M and K on Tmall.}
	\label{Table.4}
	\vspace{-20pt}
\end{table}

\subsection{Effectiveness of Compressed Model}
In our method, the model compression rate is influenced by two factors, i.e. the number of codebooks $M$ and the number of vectors in each codebook $K$. According to the compression ratio calculation equation (Eq. (11)), it is obvious that the value of $M$ has a larger influence than $K$ in theory. To investigate the effectiveness of compressed models with different compression ratios and different $M$, $K$, we conduct a comprehensive study by selecting some representative values for the two factors. To cover a wide range of compression ratios, we select $\{1, 2, 3, 4, 5\}$ for $M$ and $\{32, 128, 256, 512\}$ for $K$. We report their corresponding compression ratios and performances in Table \ref{Table.4}. From the results, we can observe that when the value of $M$ increases, the compression ratio tends to decrease, and the performances are gradually becoming better in most cases. Similarly, when $M$ is fixed, with the increase of $K$, there is a trend towards better recommendation accuracy. For the two datasets, both $M$ and $K$ are vital to the model's capacity. There is a size-accuracy trade-off when choosing different values of $M$ and $K$. 

We highlight the performance in bold when the compression ratio is 29 on Tmall where $M$ is 4 and $K$ is 32. Compared with the performance of OD-Rec where the compression ratio is 27, the highlighted performance is higher on Prec@10 and NDCG@10 and comparable on Pre@5 and NDCG@5. On the dataset of RetailRocket, we highlight the results in bold when the compression rate is 31 where $M$=3 and $K$=256. Compared with the results of OD-Rec in Table \ref{Table.3} on RetailRocket which are from the model with 30x smaller size than the uncompressed one, the highlighted performance of the new method is superior on all metrics. These observations demonstrate that when the previous method OD-Rec and the new method are compressed to the same degree, the new method has distinct advantages in recommendation accuracy, which also corroborates the effectiveness of the technical changes we have made on OD-Rec.

\begin{table}[h]
	\small
	\renewcommand\arraystretch{1.0}
	\begin{center}
		\begin{tabular}{cccccccc}
			\hline
			time &  GRU4Rec & NARM&STAMP&SR-GNN  &OD-Rec & New Student   \\ \hline
			Tmall& 0.417& 0.264& 0.175&0.310 &  0.169 & 0.021\\
			RetailRocket&0.093 & 0.096&0.094 & 0.192  &0.081 &0.012\\
			
			\hline
		\end{tabular}
	\end{center}
	\caption{Prediction time (s) on device.}
	\label{Table.5}
	\vspace{-10pt}
\end{table}

\subsection{Efficiency Analysis}
The fast local inference is an essential feature of on-device models. To demonstrate the efficiency superiority of the proposed method, we test the inference time of the new method and all the baselines including its predecessor OD-Rec on the same resource-constrained device. We first train all the models on GPU, and then they are encapsulated by PyTorch Mobile. We use Android Studio 11.0.11 to simulate virtual device environment and deploy all the models under this virtual environment to do the inference. The selected device system is Google Pixel 2 API 29. We record the prediction time for every 100 sessions of each model on the two datasets and then report the average time in Table \ref{Table.5}. Obviously, our student model is much faster than all the baselines on the two datasets. RNN-like units in GRU4Rec and NARM and the graph construction in SR-GNN leads to their prolonged inference. In particular, comparing with OD-Rec, our student model is 8x faster on Tmall and 6x faster on RetailRocket. We attribute the success to that the code matrix of items can be directly saved on the device. When doing inference, the new student model just needs to multiply it with the codebooks to get item embeddings. The code matrix consumes very limited memory because $M$ is rather small. In comparison, when using tensor-train decomposition in OD-Rec, even if the index lists can be saved, there is still a series of multiplication between indexed vectors. This efficiency analysis proves that the integration of compositional code can successfully address the computation bottleneck and reduce inference time. 

\begin{figure}[t]
	\centering
	\includegraphics[width=\textwidth]{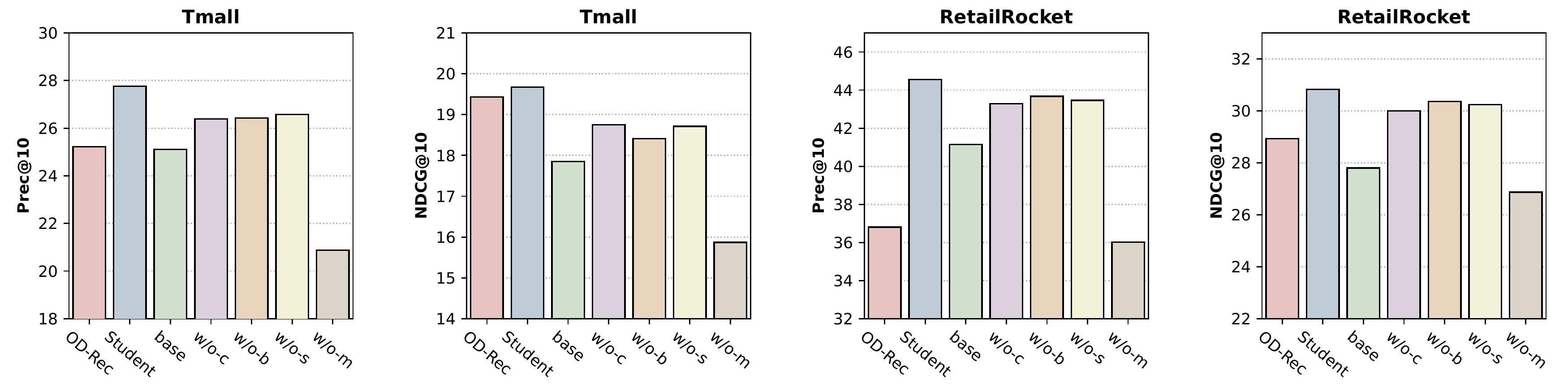}
	\caption{Ablation study.}
	\label{figure.3}
	\vspace{-10pt}
\end{figure}

\subsection{Ablation Study}
The superior performance of the new on-device model has demonstrated the effectiveness of the overall method design. To investigate the contribution of each part of the proposed method, we devise five variants, i.e. $\textbf{Stu-base}$, $\textbf{Stu-w/o-c}$, $\textbf{Stu-w/o-b}$, $\textbf{Stu-w/o-s}$ and $\textbf{Stu-w/o-m}$. $\textbf{Stu-base}$ represents the compressed model without any knowledge distillation; $\textbf{Stu-w/o-c}$ represents the one where the contrastive learning task is detached; $\textbf{Stu-w/o-b}$ means the knowledge transfer from the student to the teacher is disabled, where the parameters of the teacher model are frozen; and $\textbf{Stu-w/o-s}$ means we dispense with the soft targets distillation task; $\textbf{Stu-w/o-m}$ denotes the version where the embedding mixup proposed in Eq (16) is disabled. We present their results in Figure \ref{figure.3}. Comparing Stu-base with the full version of the student, we find that the bidirectional self-supervised knowledge distillation framework improves the model by 10.55\% on Pre@10 and 10.20\% on NDCG@10 on Tmall and 8.29\% on Pre@10 and 10.90\% on NDCG@10 on RetailRocket. The two distillation tasks and the bidirectional transfer mechanism in the knowledge distillation framework all play a part in these performance gains. Besides, it should be noted that the embedding mixup is critical. Without it, the performance drastically drops, which demonstrates that the direct knowledge transfer at the embedding level is the most effective.

\begin{figure}[t]
	\centering
	\includegraphics[width=\textwidth]{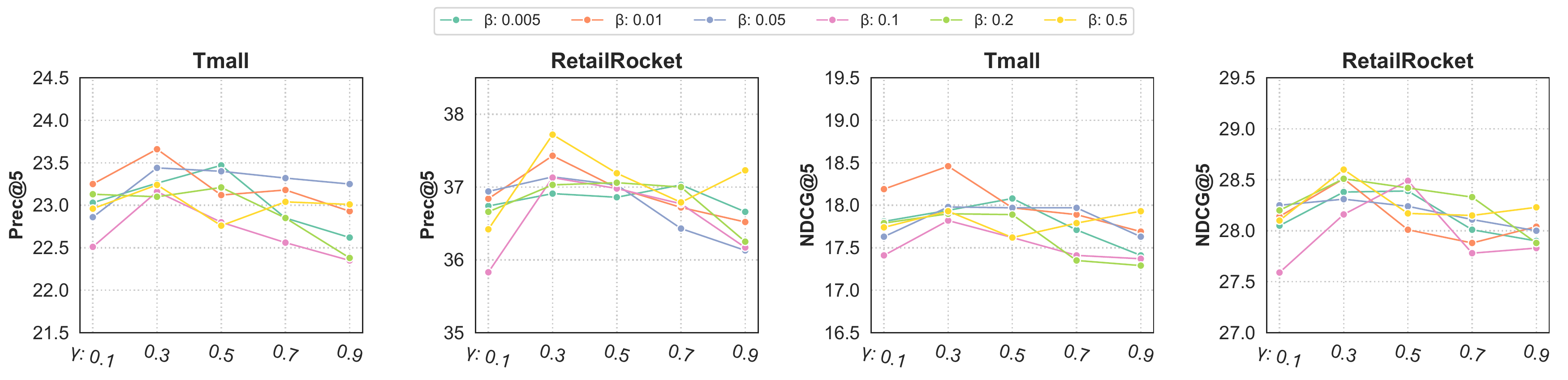}
	\caption{Influence of $M$ and $K$ on recommendation accuracy.}
	\label{figure.4}
	\vspace{-10pt}
\end{figure}

\begin{figure}[t]
	\centering
	\includegraphics[width=\textwidth]{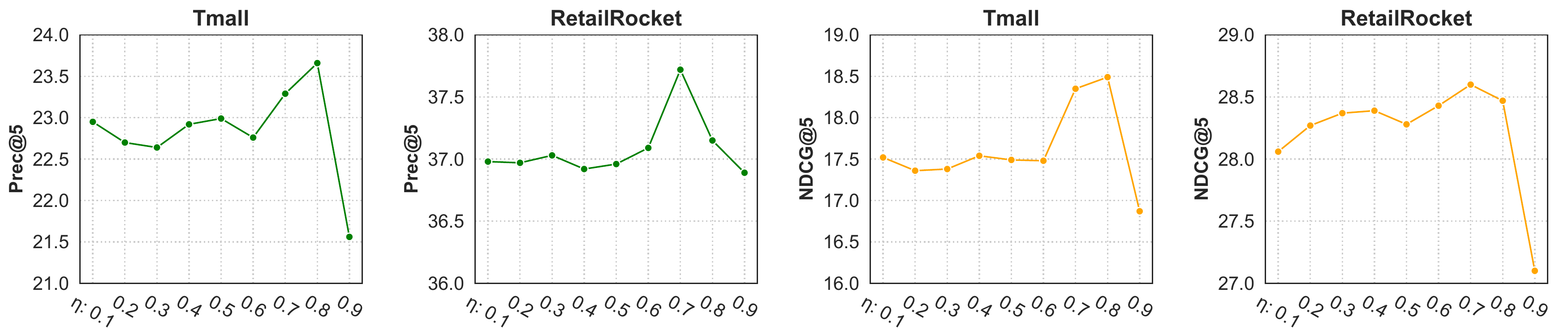}
	\caption{Influence of $\eta$ on recommendation accuracy.}
	\label{figure.5}
\end{figure}

\subsection{Hyperparameter Analysis}
In our method, there are three important hyperparameters: $\beta$, $\gamma$ and $\eta$. The first two control the effect of the two distillation tasks where $\beta$ is for the contrastive distillation task and $\gamma$ is for the soft target distillation task. $\eta\in (0,1)$ determines how much information from the teacher is used in the embedding mixup (Eq. (16)).\par
We first investigate the model's sensitivity to $\beta$ and $\gamma$ on two datasets. A large number of hyperparameter combinations are tested and we choose to report some representative ones of the two hyperparameters; they are $\{0.1, 0.3, 0.5, 0.7, 0.9\}$ for $\mathcal{\gamma}$ , $\{0.005, 0.01, 0.05, 0.1, 0.2, 0.5\}$ for $\mathcal{\beta}$. The corresponding results are presented in Fig \ref{figure.4}. When investigating the former two hyperparameters, we fix $\mathcal{\eta}$. As can be observed from Fig \ref{figure.4}, although each curve can have more than one rises and falls when tuning $\gamma$, it shows a relatively stable tendency: first goes up to the top and then gradually declines. Specifically, $\gamma$ = 0.3 is the best choice for the model on both the two datasets. Meanwhile, the best performance is reached when $\beta$=0.01 on Tmall and $\beta$=0.5 on RetailRocket. We also notice that the curves of Precision and NDCG are not always consistent in where the best performance is reached. \par
In Fig \ref{figure.5}, we report the results brought by different $\eta$, ranging from 0.1 to 0.9 with the step 0.1. When investigating $\eta$, we fix $\beta$ and $\gamma$. According to Fig. \ref{figure.5}, a larger $\eta$ can lead to better performance on both the two datasets ($\eta$=0.8 on Tmall, and $\eta$=0.7 on RetailRocket), which means a great deal of embedding information transfer from the teacher is beneficial. But when $\eta$ is 0.9, the model performance declines drastically. We conjecture that when $\eta$ is overlarge, the gradients for updating the student's parameters could be too small, resulting in underfitted parameters.


\subsection{Analysis on Distribution of Compositional Codes}
\begin{figure}[t]
	\centering
	\includegraphics[width=\textwidth]{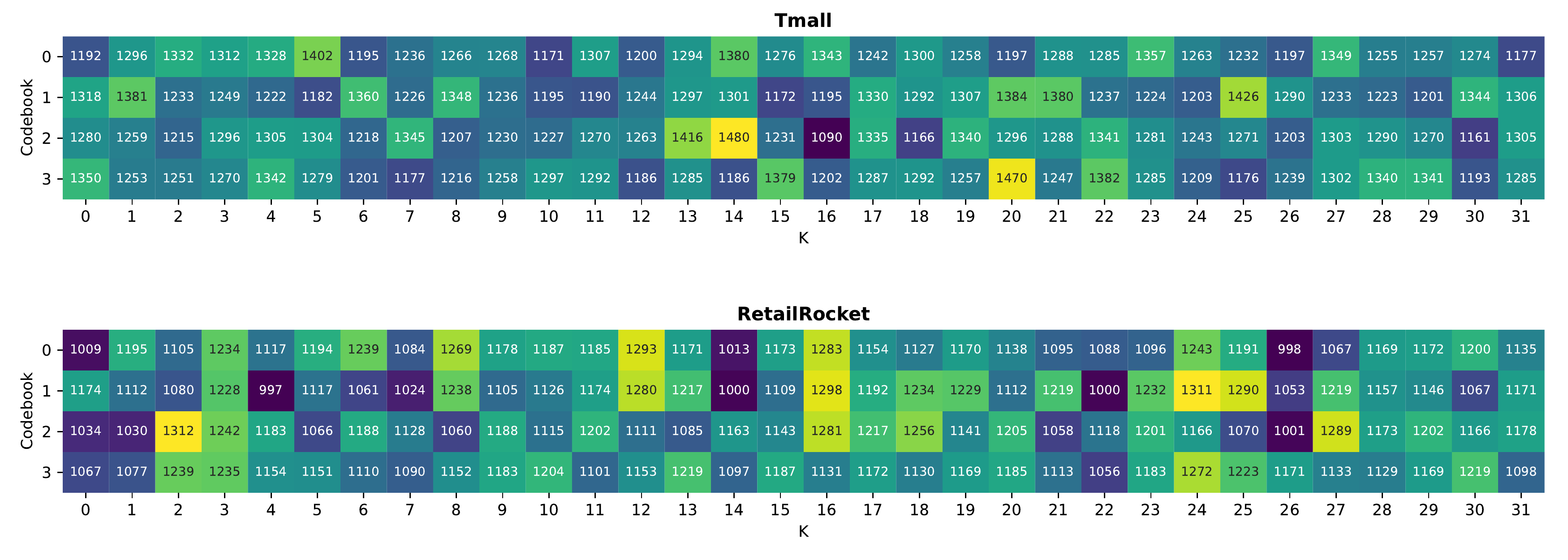}
	\caption{Heatmap of codebooks vectors on two datasets.}
	\label{figure.6}
\end{figure}

The item embedding in our on-device model is the sum of vectors looked up from $M$ codebooks by using the compositional code. To investigate the distribution of compositional codes of all items, i.e. how many times each code vector is indexed or mapped, we record the code matrix generated by the Gumbel-Softmax operation when the model reaches its best performance on two datasets. We then calculate the total times that each vector is hit and display these numbers with heatmaps in Fig. \ref{figure.6}, where each grid represents a single vector and each row represents vectors from the same codebook. For a clear and convenient display, we choose to present the distribution when $M$ = 4 and $K$ = 32 for both datasets. As shown in the figure, we can see that the distribution of compositional codes of items is relatively uniform. The numbers fall in the intervals of [1090,1480] on Tmall and [997,1312] on RetailRocket. These distributions make sense because items have different characteristics and a good encoding module should map them to different codes in order to reflect the difference. Meanwhile, some items have common characteristics and their codes should have a overlap which results in some lighter grids. Note that the values in the heatmap of Tmall are larger than those in that of RetailRocket. This is because the dataset of Tmall contains more items. The item number equals to a quarter ($\frac{1}{M}$) of the sum of the values in the heatmap.

\subsection{{Generalization Ability Analysis}}
{In this paper, our proposed model compression method and bidirectional self-supervised knowledge distillation framework are orthogonal to session-based recommendation models. To investigate the generalization ability, we conduct experiments on two other representative baseline models: NARM \cite{li2017neural} and SR-GNN \cite{wu2019session}, which are based on RNN and GNN, respectively. We present their results where we set $M$=4, $K$=32 for Tmall and $M$=3, $K$=256 for RetailRocket. NARM$^{\dagger}$ and SR-GNN$^{\dagger}$ represents the corresponding new models. Results are shown in Table 6. Compared with results in Table 3, we can directly know that NARM and SR-GNN are both enhanced, showing that our proposed method can consistently improve  a wide spectral of backbones for session-based recommendation. }
\begin{table}
	\centering
	\begin{tabular}{c|c|c|c|c|c} 
	\hline
	\makecell{Datasets}  & \makecell{Model} & \makecell{Prec@5} & \makecell{NDCG@5}  & \makecell{Prec@10}  & \makecell{NDCG@10}  \\\hline
				  &   \makecell[c]{NARM}  &  \makecell[c]{16.84}  & \makecell[c]{10.96} & \makecell[c]{19.68}  & \makecell[c]{11.66}  \\
				  Tmall	  & \makecell[c]{NARM$^{\dagger}$}  &\makecell[c]{21.19}  &  \makecell[c]{16.79}  & \makecell[c]{24.68} & \makecell[c]{17.92}     \\
      
                     &   \makecell[c]{SR-GNN}  &  \makecell[c]{19.39}  & \makecell[c]{14.29} & \makecell[c]{23.79}  & \makecell[c]{15.81}  \\
				  &   \makecell[c]{SR-GNN$^{\dagger}$}  &  \makecell[c]{21.61}  & \makecell[c]{17.14} & \makecell[c]{24.86}  & \makecell[c]{18.20}  \\\hline

				 &   \makecell[c]{NARM}  &  \makecell[c]{32.94}  & \makecell[c]{20.07} & \makecell[c]{39.18}  & \makecell[c]{21.53}  \\
				  RetialRocket	  & \makecell[c]{NARM$^{\dagger}$}  &\makecell[c]{34.93}  &  \makecell[c]{27.39}  & \makecell[c]{41.43} & \makecell[c]{29.52}     \\
      
                &   \makecell[c]{SR-GNN}  &  \makecell[c]{35.68}  & \makecell[c]{27.19} & \makecell[c]{43.31}  & \makecell[c]{29.67}  \\
				  &   \makecell[c]{SR-GNN$^{\dagger}$}  &  \makecell[c]{36.50}  & \makecell[c]{27.87} & \makecell[c]{43.99}  &  \makecell[c]{30.29} \\
	\hline
	\end{tabular}
	\caption{Performances of different baseline models.}	
	\label{Table.6}
	\vspace{-20pt}
	\end{table}

\section{Future Work and Conclusion}
In recently years, on-device session-based recommendation systems have been achieving increasing attention on account of the low energy/resource consumption and privacy protection. The constrained memory and computation resource in mobile devices has spurred a series of research for making state-of-the-art deep recommendation models fit in resource-constrained devices. Techniques such as quantization, low-rank decomposition and network pruning have been utilized to compress the regular models and achieved promising performance. However, the high compression rate often comes at the cost of slow local inference because of the cumbersome operations to recover the parameters for prediction. In this paper, we propose a new model compression method based on compositional encoding to accelerate the embedding reconstruction for faster local inference. Experimental results show that the new method can achieve 6x-8x faster inference and meanwhile exhibits better recommendation performance with a large compression ratio. 
\par 
In addition to reducing the model size to fit in mobile devices, how to regularly update on-device model is another problem. Re-training from scratch can be costly and energy-consuming. Recently, a few studies \cite{yuan2020parameter} propose the module grafting, which allows the original parameters to remain unchanged and inserts re-trained small blocks with a small number of parameters into the original model to finish model updates. This could be a promising approach for conveniently  updating on-device recommendation models. However, the major parameters to be updated in our situation are the item embedding table, which means this technique cannot be directly applied. Therefore, in our future work, we expect to target efficient model updates for on-device recommendation models.   


\bibliographystyle{ACM-Reference-Format}
\bibliography{ref}
\end{document}